\documentclass[twocolumn]{aastex631}
\usepackage[utf8]{inputenc}
\makeatletter

\providecommand{\tabularnewline}{\\}

\DeclareTextSymbolDefault{\textquotedbl}{T1}

\usepackage{booktabs}
\usepackage{float}
\usepackage{amsbsy}
\usepackage{amssymb}
\usepackage{amsmath}
\usepackage{amstext}
\usepackage{amsthm}

\usepackage{natbib}
\usepackage{graphicx}
\usepackage{array}
\usepackage{enumerate}\usepackage{bm}
\usepackage{multirow}
\usepackage{braket}
\usepackage{txfonts}
\usepackage{physics}
\usepackage[normalem]{ulem}

\def\be{\begin{equation}}
\def\ee{\end{equation}}

\makeatother

\begin{document}

\title{Secular Evolution of Quasiperiodic Eruptions: eRO-QPE2}

\author[0009-0008-5418-2977]{Zirui Zhang}

\email{zirui\_zhang@sjtu.edu.cn}

\affiliation{Tsung-Dao Lee Institute, Shanghai Jiao-Tong University, Shanghai, 520 Shengrong Road, 201210, People’s Republic of China}

\affiliation{School of Physics \& Astronomy, Shanghai Jiao-Tong University, Shanghai, 800 Dongchuan Road, 200240, People’s Republic of China}

\author[0009-0004-4654-1328]{Cong Zhou}

\affiliation{Tsung-Dao Lee Institute, Shanghai Jiao-Tong University, Shanghai, 520 Shengrong Road, 201210, People’s Republic of China}

\author[0000-0001-9608-009X]{Zhen Pan}

\email{zhpan@sjtu.edu.cn}

\affiliation{Tsung-Dao Lee Institute, Shanghai Jiao-Tong University, Shanghai, 520 Shengrong Road, 201210, People’s Republic of China}

\affiliation{School of Physics \& Astronomy, Shanghai Jiao-Tong University, Shanghai, 800 Dongchuan Road, 200240, People’s Republic of China}
\begin{abstract}
Quasi-periodic eruptions (QPEs) are recurrent soft X-ray flares from galactic nuclei. In this work, we analyze both the quiescent and eruption emission of eRO-QPE2, using 11 XMM-Newton observations spanning $4.33\,{\rm yr}$. The key findings include (1) the eruption peak count rate remains approximately constant, while the quiescent count rate shows an approximately linear decline superposed with a large-amplitude oscillation; (2) no secondary eruptions are detected between the normal eruptions and any such events would have to be fainter by more than two orders of magnitude; (3) the recurrence time decreases at an average rate of $\dot{T}_{{\rm rec}}\approx-5\times10^{-6}$. These findings are consistent with an extreme mass-ratio inspiral (EMRI) interacting with a precessing accretion disk and producing two normal eruptions per orbital period. Bayesian timing analyses further constrain the EMRI orbital eccentricity to $e<0.033$ (at $95\%$ confidence level). 
\end{abstract}

\section{Introduction}

\label{sec:intro}

Quasi-periodic eruptions (QPEs) are recurrent soft X-ray flares from galactic nuclei, with recurrence times ranging from hours to days. Since the discovery of QPEs in RX J1301.9+2747 \citep{Sun2013} and in GSN 069 \citep{miniutti_qpe_2019}, more than 10 QPE sources have been reported (see \citet{Wevers2026} for a recent review). Their remarkable properties have drawn intense interest in their physical origins and astrophysical implications. A key observational breakthrough is the detection of QPEs following the tidal disruption event (TDE) AT2019qiz which links at least some QPEs to newly formed accretion disks \citep{nicholl_tde_2024}. Together with a number of other QPE sources \citep{qref_chakraborty2025_upj,qref_baldini2026_j2344,qref_quintin2023_tormund,qref_bykov2025_tormund,qref_chakraborty2021_j0249,qref_sheng2021_gsn,qref_miniutti2023_gsn,qref_guolo2025_gsn_evolution,qref_arcodia2024_ero34,qref_zhu2025_ansky,wevers_uv_2025} where TDEs or TDE-like signatures have been identified, the TDE-QPE association has been firmly established. Recent studies further show that QPEs are preferentially found in light curves of TDEs with luminous infrared echoes and/or extending emission-line regions in the host galaxies \citep{qref_wevers2024_hosts,qref_wevers_french2024_eelr,qref_xiong2025_qiz_eelr,qref_wu2025_qiz_echo,qref_gilbert2025_morphology,qref_jiang_pan2025_embers,qref_agis2026_polarization}.

The TDE-QPE association, together with the characteristic alternating long-short pattern in QPE timing has motivated the “extreme mass-ratio inspiral (EMRI)+TDE=QPE” scenario \citep{xian_collisions_2021,linial_emri_tde_2023,franchini_impacts_2023,Tagawa:2023fpb}. In this framework, a stellar-mass black hole (BH) or a normal star crosses the TDE disk twice per orbit and shocks the gas at each passage. The shocked gas is heated and expands outward, as a result, naturally gives rise to the QPE flares. Consequently, the QPE timing encodes rich information about the EMRI systems, particularly the central SMBH mass and the EMRI orbital parameters \citep{xian_collisions_2021,zhou_probing_2024-1,zhou_probing_2024,Zhou+2025,zhou_secular_2025,Chakraborty2025,chen_qpes_2026,Zeng:2026ajf}.

However, there are still debates on the EMRI+disk model: one eruption per period vs. two eruptions per period. In the former assumption, the EMRI crosses the disk twice per orbit while only one of the passages generates a detectable flare. In the latter one, both passages generate detectable flares. These assumptions lead to different interpretations of QPE data. For GSN 069, the recurrence times alternate between $T_{{\rm long}}$ and $T_{{\rm short}}$ \citep{miniutti_qpe_2019}. Both vary with time, while their sum remains approximately constant \citep{zhou_probing_2024-1,zhou_probing_2024}. This timing pattern strongly implies that $T_{{\rm long}}+T_{{\rm short}}$ is the fundamental orbital period, with two eruptions per period.

In another stable QPE source, eRO-QPE2, the alternating long-short pattern is less persistent than in GSN 069.
Additional information is therefore needed to distinguish between the two assumptions.
Hydrodynamic simulations of BH EMRI-disk collisions show highly symmetric forward and backward ejecta, naturally favoring two eruptions per period \citep{ivanov1998,liu_quasiperiodic_2026}.
For a stellar EMRI, the one-eruption assumption identifies the $\approx  8$~ks recurrence time with the full orbital period.
Only very low-mass ($<0.14 M_\odot$) main-sequence stars can then survive the tidal field of the SMBH.
This interpretation also requires the emission from the other disk crossing to remain undetected.
We search for weaker flares between the normal eruptions and find no evidence for them, with a 95\% confidence upper limit of 0.56\% on their total counts relative to the normal eruptions.
Although star-disk collisions can produce asymmetric ejecta, much less asymmetric emission is found in simulations \citep{huang_resolving_2026}, especially for small stars allowed by the tidal constraint \citep{qref_jankovic2026_parameters}.
Debris stripped in earlier encounters can also collide with the disk and produce additional emission \citep{yao_collisions_2025,linial_debris_2025}, which is found to be comparable to the ejecta emission for eRO-QPE2 like short-period stars \citep{yao_stardisk_2026}.
The emission from disk ejecta and stellar debris therefore makes such a weak intermediate flare difficult to explain.
Together, these results strongly disfavor the one-eruption assumption.

We also investigate the evolution of the quiescent X-ray emission which roughly follows a linear decay trend superposed with a large-amplitude oscillation. This observation is compatible with a rigid disk in precession, where the accretion rate is decreasing and the disk precession modulates the observed disk emission periodically.

These non-timing data support the model of EMRI+a precessing disk generating two eruptions per period. With this model, we further perform Bayesian analyses of the eRO-QPE timing data which tightly constrain the EMRI orbital eccentricity to $e<0.033$ and the orbital period decay rate as $\dot{T}_{{\rm obt}}=-\left(1.10^{+0.07}_{-0.12}\right)\times10^{-5}$ ($95\%$ C.I.).

This paper is organized as follows. Section 2 presents the observations and light-curve analysis. Section 3 examines the timing modulations, and Section 4 presents the results of Bayesian timing inference. We summarize our results and discuss the implications in Section 5. In Appendix, we show more details of eRO-QPE2 light curves and the posteriors of EMRI+disk model parameters.  We use geometrical units $G=c=1$ unless these constants are explicitly restored.

\section{eRO-QPE2 Data}

\label{sec:data}

We analyze 11 XMM-Newton observations of eRO-QPE2 obtained between 2020 August 6 and 2024 December 5, spanning $4.33\,{\rm yr}$. The observation IDs are 0872390101, 0893810501, 0883770201, 0931791301, 0932590201, 0932590101, 0941270101, 0941270201, 0941270301, 0941270401, and 0941270501. We reprocessed the EPIC-pn observation data files using the XMM-Newton Science Analysis System (\texttt{SAS}, version 22.1.0) and the associated Current Calibration Files. All photon arrival times are barycentre-corrected in the DE405-ICRS reference system and expressed in TDB. We use the $0.2$-$2\,\mathrm{keV}$ band, bin the data at $100\,\mathrm{s}$, and fit the light curves with the phenomenological model proposed by \cite{Norris:2005ew} (see Fig.~\ref{fig:timing-fits}).

\subsection{Quiescent and Eruption Emissions}

\label{sec:emission}

\begin{figure}
\begin{centering}
\includegraphics[width=1\columnwidth,keepaspectratio]{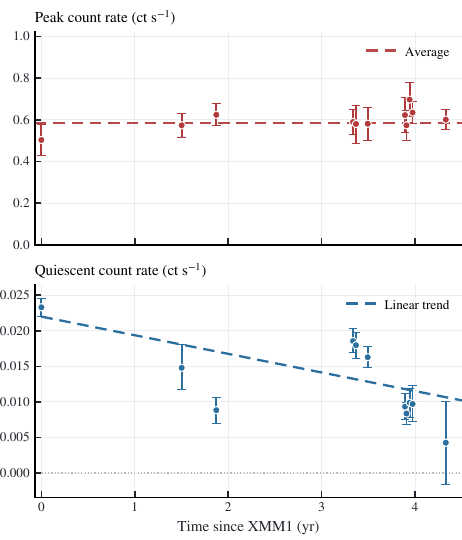} 
\par\end{centering}
\caption{Evolution of the eruption peak and quiescent count rates over the 11 XMM-Newton epochs. \textbf{Red:} eruption peak count rates and their mean. \textbf{Blue:} quiescent count rates and a linear trend.}
\label{fig:count-rates} 
\end{figure}

Fig.~\ref{fig:count-rates} shows the different long-term evolution of the two emission components. The eruption peak count rate remains approximately constant, while the quiescent count rate declines. The quiescent count rate cannot be well fitted by a linear trend alone. For example, epoch 3 is clearly below this trend.

A natural interpretation is that the two components come from different emitting regions. The quiescent X-rays may come from the inner accretion disk, as suggested by spectral modeling of eRO-QPE2 \citep{arcodia_ticking_2024,wevers_uv_2025}. In the EMRI+disk model considered here, the collisions occur at radii of  $\mathcal{O}(10^2)\,M_{\bullet}$. For comparable dimensionless viscosity and disk aspect ratio, the viscous evolution time increases with radius \citep{shakura_disk_1973}. The collision region may therefore evolve more slowly than the inner X-ray-emitting disk. This may explain why the eruption peaks remain relatively stable.

The observed quiescent count rate can also vary as the disk orientation changes. Disk precession may therefore contribute to the deviations from the overall decline \citep{arcodia_ticking_2024}. This is an observational motivation to consider a precessing disk in addition to theoretical expectations of TDE disks \citep{zanazzi_tde_2019}.

\subsection{One or Two Eruptions per Period}

\label{sec:eruption-asymmetry}

In the EMRI+disk QPE model, a stellar-mass BH or a star crosses the disk twice per orbital period. However, two disk crossings do not necessarily imply two equally detectable eruptions due to possible asymmetry in forward and backward ejecta. This leads to debates over whether there are one or two eruptions per period in QPEs. In this subsection, we will show the one-eruption assumption is strongly disfavored for eRO-QPE2.

For sBH EMRI-disk collisions, various hydrodynamic simulations show that the shocks and ejecta produced by gravitational focusing are generally symmetric on the two sides of the disk \citep{ivanov1998,liu_quasiperiodic_2026}. Two observable eruptions per orbit are therefore a natural expectation, consistent with the observations of eRO-QPE2.

For a stellar EMRI crossing a TDE disk, additional constraints come from stellar stability and the asymmetry of eruptions. 
\begin{itemize}
\item \emph{Stability criterion.} First, the star must survive the tidal field of the SMBH with $r_{{\rm Roche}}/{R_{\star}}>1$, where $R_{\star}$ is the stellar radius and $r_{{\rm Roche}}$ is the Roche lobe radius. For a star of mass $m_{\star}$, the Roche lobe radius $r_{{\rm Roche}}$ can be calculated as \citep{Paczynsky1971} 
\begin{equation}
\frac{r_{{\rm Roche}}}{r}=0.46224\left(\frac{q}{1+q}\right)^{1/3}\ ,
\end{equation}
where $q:=m_{\star}/M_{\bullet}$ is the mass ratio and $r$ is the binary separation. For a quasi-circular stellar orbit, we find 
\begin{equation}
\frac{r}{2r_{{\rm t}}}\simeq\frac{r_{{\rm Roche}}}{R_{\star}}\simeq0.4\left(\frac{T_{{\rm obt}}}{8\,{\rm ks}}\right)^{2/3}\left(\frac{m_{\star}}{M_{\odot}}\right)^{1/3}\left(\frac{R_{\star}}{R_{\odot}}\right)^{-1}\ .\label{eq:tidal-period}
\end{equation}
where $r_{{\rm t}}=R_{\star}(M_{\bullet}/m_{\star})^{1/3}$ is the tidal radius. Under the one-eruption assumption for eRO-QPE2, $T_{{\rm obt}}\simeq T_{{\rm rec}}\approx8~\mathrm{ks}$. Using the stellar mass-radius relation $R_{\star}\propto m^{0.8}_{\star}$, we find that the stability criterion excludes all main-sequence stars with $m_{\star}>0.14M_{\odot}$. 
\item \emph{Secondary flares in eRO-QPE2 data.} The one-eruption assumption requires highly asymmetric forward and backward ejecta or stellar debris, with the stronger producing the observed eruption and the weaker producing a much fainter one. We therefore search for secondary flares in eRO-QPE2 by stacking all 11 XMM-Newton observations. We align the flare peak times and stack the data bins from one flare start to the next. The upper panel of Fig.~\ref{fig:stacked-light-curve} shows the resulting stacked light curve, with no obvious secondary flare in the quiescent interval. We fit the quiescent segment from $3~\mathrm{ks}$ to the end of the stack, using the same phenomenological model as for the normal eruption. We find no evidence for a secondary flare and obtain a 95\% confidence upper limit of $N_{{\rm small}}/N_{{\rm large}}<0.56\%$. Here $N_{{\rm small}}$ and $N_{{\rm large}}$ are the total counts of the hypothesized faint flare and the normal eruption in the stacked light curve, respectively. 
\begin{figure}
\includegraphics[width=1\columnwidth,keepaspectratio]{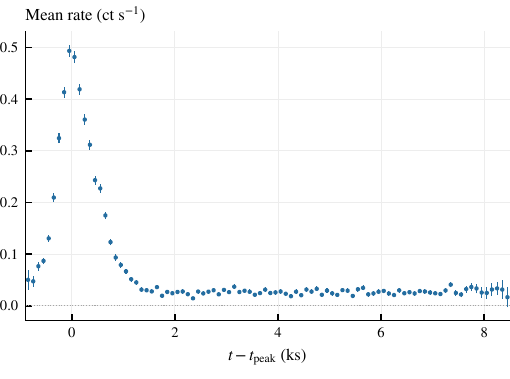} \includegraphics[width=1\columnwidth,keepaspectratio]{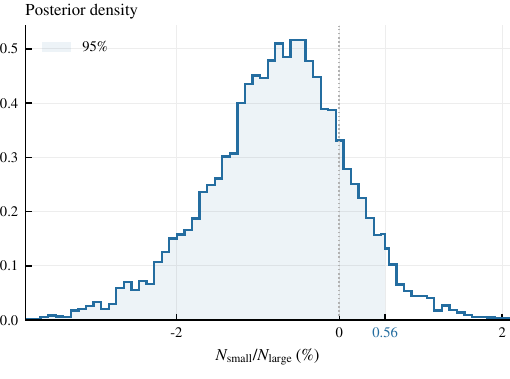} \caption{Upper panel: Stacked eruption-quiescent signal of all 11 XMM epochs. Lower panel: Posterior density of $N_{{\rm small}}/N_{{\rm large}}$ from the Bayesian analysis of stacked signal, where $N_{{\rm small}}$ and $N_{{\rm large}}$ are the total counts of the hypothesized secondary flare and the normal flare, respectively. }
\label{fig:stacked-light-curve} 
\end{figure}
\item \emph{Eruption asymmetry in simulations}. A number of recent simulation studies of star-disk collisions reveal asymmetry in the forward and backward ejecta, though the degree of asymmetry found in different studies varies depending on the setup details of simulations \citep{liu_quasiperiodic_2026,qref_huang2025_multiband,huang_resolving_2026,qref_jankovic2026_rhd,qref_jankovic2026_parameters,qref_zhu2026_atmosphere}. As shown by \citet{huang_resolving_2026}, both numerical resolution and 3D geometry in modeling star--disk collisions are key in quantifying the eruption energy and eruption geometry. High-resolution simulations of oblique impacts show that including disk rotation allows for easier shock breakout from the disk’s backside and substantially reduces the luminosity contrast between forward and backward ejecta. For a face-on observer, their photospheric estimates give a peak bolometric luminosity ratio of $\gtrsim50$ without disk rotation and $\sim10$ with rotation. At later times, these ratios are $\sim5$ and $\lesssim2$, respectively. Recent 3D radiation-hydrodynamic simulations \citep{qref_jankovic2026_parameters} of star-disk collisions show that the emission energy of backward ejecta is comparable to and sometimes higher than the forward counterpart, with a robust lower limit $E_{{\rm back}}/E_{{\rm for}}>3\%$, across the parameter space explored. The asymmetry largely depends on the star size with an approximate scaling relation $E_{{\rm back}}/E_{{\rm for}}\propto(R_{\star}/R_{\odot})^{-0.8}$. The strongest asymmetry occurs for the biggest stars with $R_{\star}=4R_{\odot}$, and reduces to $E_{{\rm back}}/E_{{\rm for}}\simeq0.18$ for small stars with $R_{\star}=0.5R_{\odot}$. In the case of eRO-QPE2, more symmetric forward and backward ejecta emissions are expected considering the stability requirement $R_{\star}<0.2R_{\odot}$. These simulation results disfavor the one-eruption assumption for eRO-QPE2.

Repeated star-disk collisions also strip material from the stellar envelope. The stellar debris can subsequently collide with the disk and produce additional emission \citep{yao_collisions_2025,linial_debris_2025}. Recent simulations \citep{yao_stardisk_2026} suggest that the relative importance of ejecta and stellar debris depends on the orbital period or the ratio $r/r_{{\rm t}}$. At longer orbital periods (large $r/r_{{\rm t}}$), shocked stellar debris dominates. At shorter periods, the ejecta contribution becomes stronger. In their shortest-period simulation, with $r/r_{{\rm t}}=3.5$, the shocked disk and stellar-debris masses become comparable. For eRO-QPE2, the shortest-period QPE source with $r/r_{{\rm t}}\simeq2-2.3$ for $m_{\star}=0.1-0.14M_{\odot}$, the stellar debris is expected to contribute comparable emission. These simulation results leave no room for the one-eruption assumption in the case of eRO-QPE2. 
\end{itemize}

As a result, the one-eruption interpretation of eRO-QPE2 requires a rather low-mass ($m_{\star}<0.14M_{\odot}$) star to survive at the smaller orbital radius. It also requires the intermediate emission from the other collision, including any contribution from stellar debris, to remain below our detection limit, $N_{{\rm small}}/N_{{\rm large}}<0.56\%$. The backward ejecta found in the simulations alone makes such a weak intermediate signal less easy to explain. The additional emission from stellar debris found in the simulations makes it even harder to explain. Therefore existing interpretations of eRO-QPE2 with the one eruption per period assumption are ungrounded. We consider two normal eruptions of comparable strength per orbit in the timing analysis below.

\section{Modulations in QPE Timing}

\label{sec:modulation}

We can characterize QPE flare timing modulations through the recurrence time between adjacent flares, $T_{n}=t_{n+1}-t_{n}$. This section describes several basic modulation mechanisms in the EMRI+disk model and how they, separately or in combination, appear in the recurrence times.

\begin{figure}
\begin{centering}
\includegraphics[width=1\columnwidth,keepaspectratio]{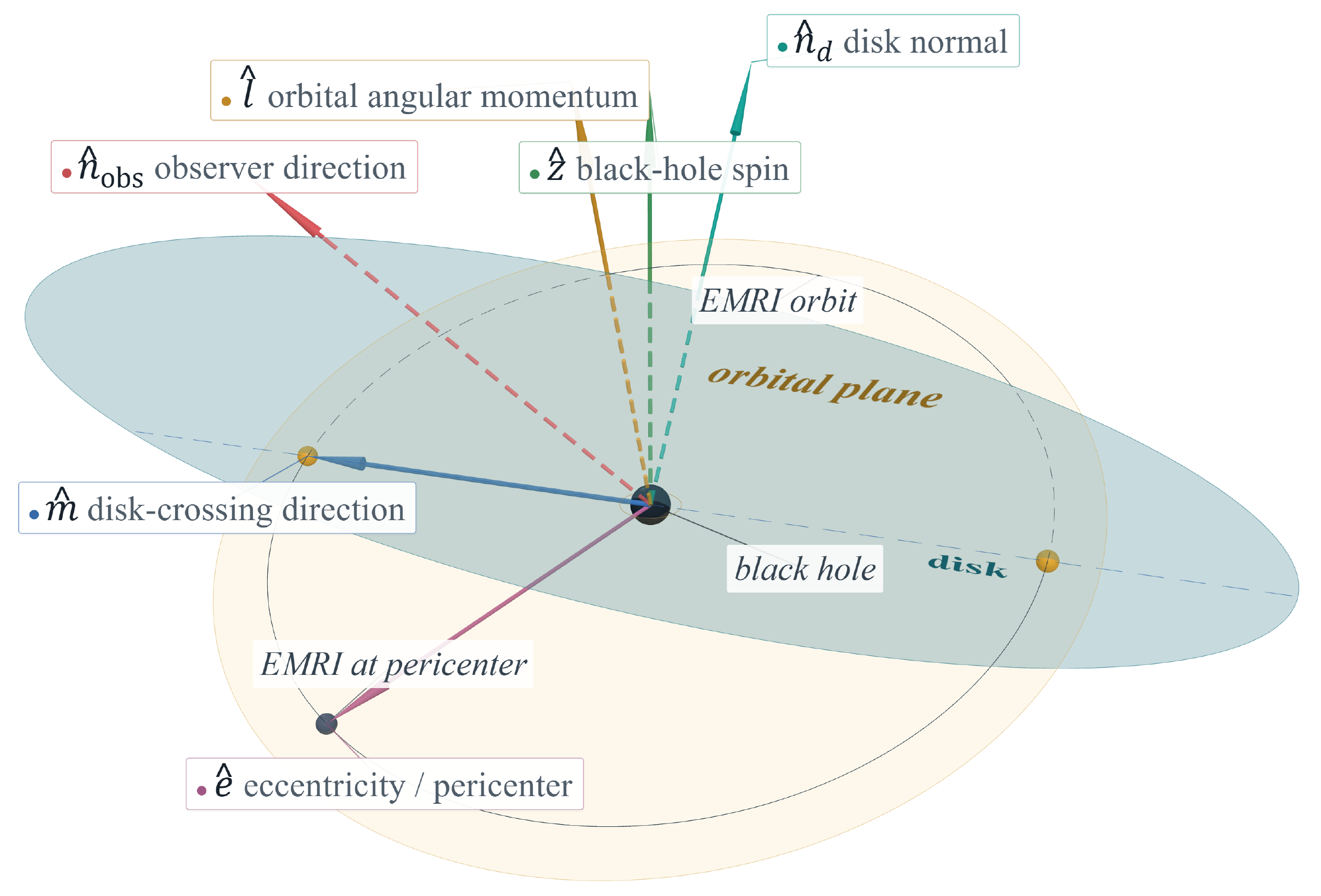} 
\par\end{centering}
\caption{Geometric configuration and vector definitions for the EMRI+disk system.}
\label{fig:geometry} 
\end{figure}

For clarity, we define the following unit vectors: 
\begin{itemize}
\item $\hat{l}$: direction of the EMRI orbital angular momentum; 
\item $\hat{\boldsymbol{d}}$: disk normal vector; 
\item $\hat{\boldsymbol{z}}$: black hole spin direction; 
\item $\hat{\boldsymbol{n}}_{\mathrm{obs}}$: direction from the black hole to the observer; 
\item $\hat{e}$: direction of the eccentricity vector, from the black hole to the EMRI orbit's pericenter; 
\item $\hat{\boldsymbol{m}}=\left(\hat{\boldsymbol{l}}\times\hat{\boldsymbol{d}}\right)/\left|\hat{\boldsymbol{l}}\times\hat{\boldsymbol{d}}\right|$: direction from the black hole to the disk crossing point when the effective disk thickness is $0$. 
\end{itemize}
Fig.\,\ref{fig:geometry} illustrates these definitions. We denote the angle between the disk normal and the SMBH spin by $\beta$, and the angle between the orbital plane normal and the spin by $\iota$. The modulations discussed below can then be divided into two topological branches, $\beta>\iota$ and $\beta<\iota$, separated by $\beta=\iota$.

\subsection{Intrinsic Modulation}

Intrinsic modulation in QPE timing arises from intrinsic dynamics of a Kerr orbit, including modulations induced by apsidal and L-T precession. Apsidal precession of an eccentric EMRI with orbital eccentricity $e$ produces anti-phase modulations in the recurrence times at even and odd timing indices, which under an equatorial disk assumption have the form like 
\begin{equation}
\begin{aligned}T_{{\rm even}}(t) & \simeq C+A_{\mathrm{aps}}\sin(\omega_{{\rm aps}}t+\phi_{{\rm aps}})\\
T_{{\rm odd}}(t) & \simeq C-A_{\mathrm{aps}}\sin(\omega_{{\rm aps}}t+\phi_{{\rm aps}})\,,
\end{aligned}
\label{eq:apsidal-modulation}
\end{equation}
with $C=T_{{\rm obt}}/2$. We can estimate the amplitude as 
\begin{equation}\label{eq:apsidal-amplitude}
A_{{\rm aps}}\simeq\frac{2e}{\pi}T_{{\rm obt}}\simeq108.2\,{\rm s}\left(\frac{e}{0.01}\right)\left(\frac{T_{{\rm obt}}}{17\,{\rm ks}}\right)\,.
\end{equation}
This describes the well-known alternating long-short pattern observed in GSN 069. For L-T precession in non-grazing condition, if an equatorial disk is considered, its contribution to the modulation is $A_{{\rm LT}}=0$. In general, $A_{{\rm LT}}\lesssim T^{2}_{{\rm obt}}/2T_{{\rm LT}}\ll A_{{\rm aps}}$, which is far below the observational uncertainties and systematic errors and is negligible.

\subsection{External Modulation}

QPE timing can also be modulated by factors external to the EMRI orbit, such as disk precession or orbital motion induced by another SMBH. For a precessing disk with a circular EMRI orbit, the recurrence-time modulations at even and odd indices are then in-phase: 
\begin{equation}
T_{{\rm even}}(t)=T_{{\rm odd}}(t)\simeq C+A_{\mathrm{disk}}\sin(\omega_{{\rm disk}}t+\phi_{{\rm disk}})\ ,\label{eq:disk-precession}
\end{equation}
where 
\begin{equation}
C=\begin{cases}
\frac{T_{{\rm obt}}}{2} & \beta<\iota\\
\frac{\pi}{\Omega_{{\rm obt}}\mp\omega_{{\rm disk}}}\simeq\frac{T_{{\rm obt}}}{2}\pm\frac{T^{2}_{{\rm obt}}}{2\tau_{{\rm p}}} & \beta>\iota
\end{cases}\label{eq:mean-recurrence-time}
\end{equation}
the sign in Eq.~\eqref{eq:mean-recurrence-time} corresponds to prograde and retrograde precession of the disk. An interesting implication of Eq.~\eqref{eq:mean-recurrence-time} is the apparent period $C$ may increase with time due to the decrease of the disk precession rate $\omega_{{\rm disk}}$ as already noted by \citet{zhou_probing_2024-1}. This simple geometric effect may be responsible for the apparent period growth in the QPEs of ZTF19acnskyy \citep{Chakraborty:2026opq}.

\begin{figure}
\begin{centering}
\includegraphics[width=1\columnwidth,keepaspectratio]{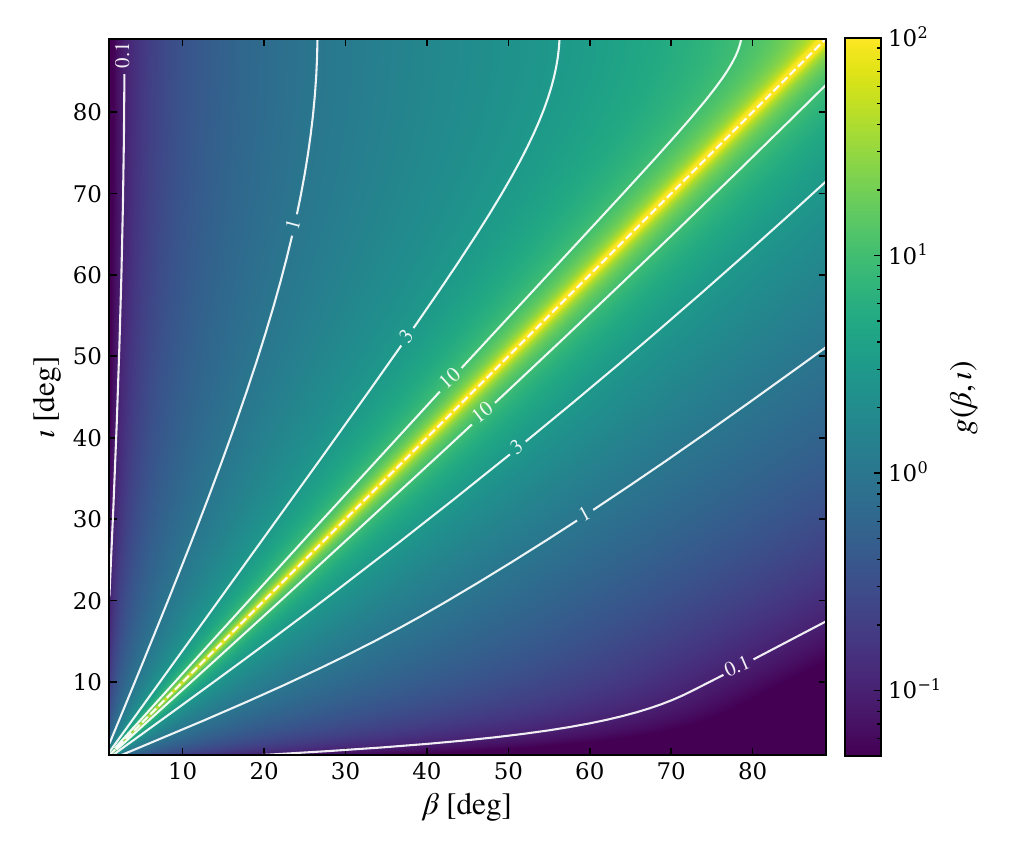} 
\par\end{centering}
\caption{The geometric factor $g(\beta,\iota)$ as a function of the disk and orbital inclinations.}
\label{fig:geometric-factor} 
\end{figure}

The amplitude of the oscillation $A_{\mathrm{ext}}$ can be estimated as 
\begin{equation}
A_{\mathrm{disk}}\simeq27.9\,{\rm s}\left(\frac{T_{{\rm obt}}}{17\,{\rm ks}}\right)^{2}\left(\frac{\tau_{{\rm p}}}{30\,{\rm d}}\right)^{-1}g(\beta,\iota)\:,\label{eq:disk-modulation-amplitude}
\end{equation}
while $g(\beta,\iota)$ is a relatively complex function describing the dependence on the disk and orbital inclinations, as shown in Fig.~\ref{fig:geometric-factor}. Over most of the parameter space, $g$ is of order $0.1$ to $10$. Near $|\beta-\iota|\to0$, the grazing geometry causes $g$ to diverge and $A_{{\rm disk}}$ to increase sharply, producing a prominent timing signal. However, since such a signal has not been found in current QPE signals, we can still regard $g$ as a function below $10$ when considering such kind of modulation. 

The orbital motion around a companion SMBH introduces a negligible change on EMRI disk crossing times. Similar to L-T precession, it makes a difference to observed QPE timing only when combined with geometry of photon paths.

\subsection{Geometric Effects}

The geometry of the EMRI+disk system can also affect the recurrence times, mainly through light-travel-time differences between the two flares in each orbit cycle and the effective disk thickness at the crossing points. The recurrence times then alternate between two constant values, $T_{{\rm long}}$ and $T_{{\rm short}}$, in orbital timescale, with recurrence time at even and odd indices behaving like: 
\begin{equation}
T_{{\rm even}}=C_{1},\ T_{{\rm odd}}=C_{2}\ ,\label{eq:geometric-modulation}
\end{equation}
rather than $T_{{\rm even}}=T_{{\rm odd}}=C$. Static geometric modulation can be regarded as intrinsic modulation except with zero frequency.

If only the effective disk half-thickness $H$ is considered, we have: 
\begin{equation}
\begin{aligned}|C_{1}-C_{2}|_{H} & \simeq\frac{2T_{{\rm obt}}}{\pi}\arcsin\left[\frac{H}{p|\hat{\boldsymbol{l}}\times\hat{\boldsymbol{d}}|}\right]\\
 & \simeq\frac{21.6\,{\rm s}}{|\hat{\boldsymbol{l}}\times\hat{\boldsymbol{d}}|}\,\left(\frac{H}{M_{\bullet}}\right)\left(\frac{T_{{\rm obt}}}{17\,{\rm ks}}\right)\left(\frac{p}{500M_{\bullet}}\right)^{-1}\frac{\arcsin x}{x}\,,
\end{aligned}
\end{equation}
where we have defined a quantity $x$ that is small in the non-grazing case with the grazing correction term $\arcsin x/x$, and $H$ here is the effective half-thickness of the disk. If only the light-travel-time is considered, 
\begin{equation}
\begin{aligned}|C_{1}-C_{2}|_{{\rm path}} & \simeq\frac{4p}{c}|\hat{\boldsymbol{m}}\cdot\hat{\boldsymbol{n}}_{{\rm obs}}|\\
 & \simeq484\,{\rm s}\left|\hat{\boldsymbol{m}}\cdot\hat{\boldsymbol{n}}_{{\rm obs}}\right|\left(\frac{T_{{\rm obt}}}{17\,{\rm ks}}\right)\left(\frac{p}{500M_{\bullet}}\right)^{-1/2}\ .
\end{aligned}
\end{equation}

As the EMRI+disk system evolves, the angular factors $|\hat{l}\times\hat{d}|$ and $|\hat{m}\cdot\hat{n}_{{\rm obs}}|$ in the two expressions above also evolve. Through these factors, geometric effects couple to intrinsic and external modulation, producing more complicated modulation patterns.

\subsection{Geometric + Intrinsic Modulation}

We now combine intrinsic modulation with geometric effects. Under non-grazing conditions, disk thickness mainly introduces constant shifts in the recurrence times at even and odd indices. Combining these shifts with apsidal precession gives 
\begin{equation}
\begin{aligned}T_{{\rm even}}(t) & \simeq C_{1}+A_{\mathrm{aps}}\sin(\omega_{{\rm aps}}t+\phi_{{\rm aps}})\\
T_{{\rm odd}}(t) & \simeq C_{2}-A_{\mathrm{aps}}\sin(\omega_{{\rm aps}}t+\phi_{{\rm aps}})\ .
\end{aligned}
\label{eq:geometric-apsidal-modulation}
\end{equation}
Since L-T precession does not modulate disk crossing times, Eq.~\eqref{eq:geometric-modulation} still holds.

We next consider how light-travel-time effects couple to intrinsic modulation. Apsidal precession changes the geometry only weakly, with a negligible modulation amplitude of $A_{\rm geom}^{\rm aps}\lesssim pe^2/c\ll A_{\rm aps}$. In contrast, L-T precession introduces a non-negligible anti-phase modulation on the L-T timescale: \begin{equation}
\begin{aligned}T_{{\rm even}}(t) & \simeq C+A_{\mathrm{geom}}\sin(\omega_{{\rm LT}}t+\phi_{{\rm LT}})\\
T_{{\rm odd}}(t) & \simeq C-A_{\mathrm{geom}}\sin(\omega_{{\rm LT}}t+\phi_{{\rm LT}})\ .
\end{aligned}
\end{equation}
For an equatorial disk, we estimate 
\begin{equation}
A_{{\rm geom}}\simeq\frac{2p}{c}\sin\theta_{{\rm obs}}\simeq242\,{\rm s}\left(\frac{T_{{\rm obt}}}{17\,{\rm ks}}\right)\left(\frac{p}{500M_{\bullet}}\right)^{-1/2}\sin\theta_{{\rm obs}}\ .
\end{equation}
Here $\theta_{{\rm obs}}$ is the angle between the observer direction and the SMBH spin direction.

\subsection{Geometric + External Modulation}

Similarly, combining external modulation with a finite disk thickness under non-grazing conditions gives 
\begin{equation}
\begin{aligned}T_{{\rm even}}(t) & \simeq C_{1}+A_{\mathrm{disk}}\sin(\omega_{{\rm disk}}t+\phi_{{\rm disk}})\\
T_{{\rm odd}}(t) & \simeq C_{2}+A_{\mathrm{disk}}\sin(\omega_{{\rm disk}}t+\phi_{{\rm disk}})\ ,
\end{aligned}
\label{eq:geometric-external-modulation}
\end{equation}
which is straightforward.

Precession of a non-equatorial disk coupling to the viewing angle produces an anti-phase modulation 
\begin{equation}
\begin{aligned}T_{{\rm even}}(t) & \simeq C\\
 & +A_{\mathrm{disk}}\sin(\omega_{{\rm disk}}t+\phi_{{\rm disk}})+A_{{\rm geom}}\sin(\omega_{{\rm disk}}t+\phi^{{\rm disk}}_{{\rm geom}})\\
T_{{\rm odd}}(t) & \simeq C\\
 & +A_{\mathrm{disk}}\sin(\omega_{{\rm disk}}t+\phi_{{\rm disk}})-A_{{\rm geom}}\sin(\omega_{{\rm disk}}t+\phi^{{\rm disk}}_{{\rm geom}})\ .
\end{aligned}
\end{equation}
For an EMRI orbit in the equatorial plane, we have the same $A_{{\rm geom}}$ as in the L-T case.

The orbital motion around a companion SMBH does not change the EMRI disk-crossing times. Both even and odd eruptions share the additional light-travel-time delay, giving an in-phase modulation: 
\begin{equation}
\begin{aligned}T_{{\rm even}}(t) & \simeq C+A_{{\rm SMBHB}}\sin(\omega_{{\rm SMBHB}}t+\phi_{{\rm SMBHB}})\\
T_{{\rm odd}}(t) & \simeq C+A_{{\rm SMBHB}}\sin(\omega_{{\rm SMBHB}}t+\phi_{{\rm SMBHB}})\ .
\end{aligned}
\end{equation}

~\\
 To summarize, intrinsic apsidal precession introduces an anti-phase modulation in the even and odd recurrence times, while external effects such as disk precession and orbital motion around a second SMBH introduce in-phase modulations. For a fixed geometry, light-travel-time delays also give an anti-phase contribution, but at zero frequency. In a general EMRI+disk model, the combined modulation is more complex than these individual contributions. For example, disk precession changes the apsidal-modulation frequency to $|\omega_{{\rm aps}}-\omega_{{\rm disk}}|$ when $\beta>\iota$, without changing the amplitude estimate or its anti-phase character. The simple analytic expressions above can not fully describe these combined effects. We therefore use osculating Kerr orbits together with the actual disk-crossing geometry for the timing analysis \citep{zhou_secular_2025}. The estimates in this section provide physical intuition for interpreting the fitting results.

\section{QPE Timing analysis: EMRI+disk}

\label{sec:timing}

\subsection{Does the geodesic assumption work?}

The EMRI orbital period decays with time as a result of EMRI-disk collisions. Over a sufficiently short interval satisfying 
\begin{equation}
\frac{N^{2}}{2}\left|\dot{T}_{{\rm rec}}\right|T_{{\rm rec}}\lesssim\sigma(t_{0})\ ,
\end{equation}
the geodesic assumption $\dot{T}_{{\rm rec}}=0$ may still be a good approximation, where $\sigma(t_{0})$ is the measurement uncertainty of QPE timing. Here $N$ is the number of eruptions within the interval $t_{{\rm span}}$, and $T_{{\rm rec}}$ and $\dot{T}_{{\rm rec}}$ are the mean recurrence time and its change rate. The corresponding validity condition is 
\begin{equation}
t_{{\rm span}}=NT_{{\rm rec}}<6.5\,{\rm d}\left(\frac{\sigma}{100\,{\rm s}}\right)^{1/2}\left(\frac{T_{{\rm rec}}}{8\,{\rm ks}}\right)^{1/2}\left|\frac{\dot{T}_{{\rm rec}}}{5\times10^{-6}}\right|^{-1/2}\ .
\end{equation}
Over longer intervals, the geodesic assumption fails and leads to numerical artifacts when fitting QPE timing data with the EMRI+disk model \citep[see][for an explicit example]{Zhou2026}. A forced geodesic model is then required for the timing analysis \citep{zhou_secular_2025}.

\subsection{Bayesian timing inference}

\label{sec:inference}

A Bayesian framework for inferring EMRI dynamics from QPE timing data has been established \citep{zhou_probing_2024,zhou_probing_2024-1,Zhou+2025,zhou_secular_2025}. In the EMRI+disk model, the observed QPE arrival time $t_{{\rm obs}}$ is written as 
\begin{equation}
t_{{\rm obs}}=t_{{\rm crs}}+\delta t_{{\rm geo}}+\delta t_{{\rm shap}}\ ,
\end{equation}
where $t_{{\rm crs}}$ is the EMRI disk-crossing time, $\delta t_{{\rm geo}}$ and $\delta t_{{\rm shap}}$ are the geometric and Shapiro delays of photons propagating from the emission location to the observer, respectively. Nine parameters specify the EMRI trajectory in Kerr spacetime: the orbital eccentricity $e$, the polar turning point $\theta_{\min}$, the three initial phases $q_{r,0}$, $q_{z,0}$ and $q_{\phi,0}$, the initial orbital period $T_{{\rm obt}}$, its derivative $\dot{T}_{{\rm obt}}$ due to EMRI-disk collisions, and the SMBH mass $M_{\bullet}$ and spin $a$. The semilatus rectum $p$ is determined by $T_{{\rm obt}}$, $M_{\bullet}$ and $e$. Three additional parameters specify the disk orientation and precession: the disk normal inclination $\beta_{0}$, the initial phase in the disk $\phi$ direction $\alpha_{0}$, and the disk precession period $\tau_{{\rm p}}.$ The observer's polar angle $\theta_{{\rm obs}}$ is also required to calculate the photon delays.

We use a Gaussian likelihood to compare the observed flare starting times $\boldsymbol{d}=\{t_{k}\}$ with the model predictions $t^{{\rm mod}}_{N_{k}}(\boldsymbol{\Theta})$. The measurement uncertainties $\sigma_{k}$ and an additional timing systematics $\sigma_{{\rm sys}}$ are combined as $s^{2}_{k}=\sigma^{2}_{k}+\sigma^{2}_{{\rm sys}}$. Since the number of cycles across the observing gaps is unknown, we adopt a uniform prior on the cycle numbers $\boldsymbol{N}_{{\rm cyc}}=\{N_{k}\}$ and marginalize over them, 
\begin{equation}\label{eq:likelihood}
\mathcal{L}_{{\rm timing}}(\boldsymbol{d}\mid\boldsymbol{\Theta})\propto\sum_{\boldsymbol{N}_{{\rm cyc}}\in\mathcal{C}}\prod_{k}\frac{1}{\sqrt{2\pi}\,s_{k}}\exp\!\left[-\frac{\left[t_{k}-t^{{\rm mod}}_{N_{k}}(\boldsymbol{\Theta})\right]^{2}}{2s^{2}_{k}}\right]\ .
\end{equation}
Here $\Theta$ denotes the model parameters. The index $N_{k}$ identifies the predicted disk crossing matched to the $k$-th observed flare, and $\mathcal{C}$ contains all allowed matching assignments within the search windows.

The priors used are summarized in Table \ref{tab:timing-priors}. We sample the posteriors with the \texttt{dynesty} nested sampler implemented in \texttt{bilby}, using $N_{{\rm live}}=1000$ live points and stopping when the estimated remaining change in the log evidence satisfies $\Delta\ln Z<0.1$.

\begin{table}
\caption{Priors used for the Bayesian inference of eRO-QPE2.}
\label{tab:timing-priors} 
\begin{centering}
\begin{tabular}{cc}
\toprule 
$\Theta$  & $\pi(\Theta)$\tabularnewline
\midrule 
$T_{{\rm obt}}\,[{\rm ks}]$  & $\mathcal{U}(14.4,25)$\tabularnewline
$e$  & $\mathcal{U}(0,0.9)$\tabularnewline
$\sin\theta_{\min}$  & $\mathcal{U}(0,1)$\tabularnewline
$q_{r,0}$  & $\mathcal{U}(0,2\pi)$\tabularnewline
$q_{z,0}$  & $\mathcal{U}(0,2\pi)$\tabularnewline
$q_{\phi,0}$  & $\mathcal{U}(0,2\pi)$\tabularnewline
$\log_{10}(M_{\bullet}/M_{\odot})$  & $\mathcal{U}(4,7)$\tabularnewline
$a$  & $\mathcal{U}(0,1)$\tabularnewline
$\cos\theta_{{\rm obs}}$  & $\mathcal{U}(0,1)$\tabularnewline
$\alpha_{0}$  & $\mathcal{U}(0,2\pi)$\tabularnewline
$\tau_{p}\,[{\rm day}]$  & $\mathcal{U}(0.5,100)$\tabularnewline
$\cos\beta_{0}$  & $\mathcal{U}(0,1)$\tabularnewline
$|\dot{T}_{{\rm obt}}|\,[{\rm s\,s^{-1}}]$  & $\mathcal{U}(0,10^{-4})$\tabularnewline
$\sigma_{{\rm sys}}\,[{\rm ks}]$  & $\mathcal{U}(0,2)$\tabularnewline
\bottomrule
\end{tabular}
\par\end{centering}
Note. $\mathcal{U}(A,B)$ denotes a uniform distribution on $[A,B]$. The angular priors are uniform in the quantities listed in the table to ensure that the priors are truly isotropic in space, with $0\leq\cos\theta_{\min}\leq1$ and $0\leq\theta_{{\rm obs}},\beta_{0}\leq\pi/2$. Periodic boundaries are used for phase parameters $q_{r,0}$, $q_{z,0}$, $q_{\phi,0}$, and $\alpha_{0}$. 
\end{table}

\subsection{Results}

\label{sec:timing-results}

The disk parameters may have evolved over the more than four years covered by XMM~1-11. We therefore fit epochs 1-4 and 4-11 independently and examine the consistency of the two fits (Fits 1 and 2 hereafter).

The posteriors of model parameters are shown in Fig.~\ref{fig:corner}. The initial orbital period is the best-constrained parameter with $95\%$ confidence constraints
\begin{equation}
\begin{aligned}T_{{\rm obt}}(t_{{\rm ref},1}=59067.89) & =17.41^{+0.05}_{-0.06}\ \,{\rm ks}\quad({\rm Fit\ 1})\ ,\\
T_{{\rm obt}}(t_{{\rm ref},4}=60286.89) & =16.39^{+0.15}_{-0.13}\ \,{\rm ks}\quad({\rm Fit\ 2})\ .
\end{aligned}
\end{equation}
Here $t_{{\rm ref},1}$ and $t_{{\rm ref},4}$ are the starting times (MJD, TDB) of the first observed flares in epochs 1 and 4, respectively. Combining the two period measurements over the $3.3\,{\rm yr}$ between the reference epochs gives a mean orbital decay rate $\langle\dot{T}_{{\rm obt}}\rangle\approx-0.97\times10^{-5}$, close to the Fit 1 constraint $\dot{T}_{{\rm obt}}=-\left(1.10^{+0.07}_{-0.12}\right)\times10^{-5}$.  The orbital decay rate from Fit 2 is also consistent but less well constrained with $\dot{T}_{{\rm obt}}=-\left(0.40^{+0.62}_{-0.12}\right)\times10^{-5}$. These constraints are consistent with previous analyses based on a subset of data \citep{arcodia_ticking_2024,pasham_alive_2024,zhou_secular_2025}.
In fact, $\dot{T}_{{\rm obt}}$ of epochs 4-11 can also be inferred from the stable eruption emission count rate across all 11 epochs (see Fig.~\ref{fig:count-rates}), which implies a stable orbital decay rate. This inference can be tested with follow-up observations spanning a longer time interval.

Both fits favor a low-eccentricity orbit, with 95\% confidence upper limits $e<0.026$ and $e<0.033$ for Fits 1 and 2, respectively. This low-eccentricity measurement is well expected from the weak anti-phase modulation in the timing of eRO-QPE2. As a result, the EMRI apsidal precession period, the orbital size and the SMBH mass are not tightly constrained. The disk model parameters are weakly constrained and their posteriors of the two fits are broadly consistent, revealing no clear evidence for evolution of disk parameters over the 4 years. 

The best-fitting timing models are shown in the lower panels of Fig.~\ref{fig:timing-fits}. The long-short contrast, $T_{{\rm long}}-T_{{\rm short}}$, is approximately $650\,{\rm s}$ in epoch 1 and 3, but decreases to about $100\,{\rm s}$ in later epochs. In the best-fitting configurations, geometric light-travel-time effects contribute up to approximately $500\,{\rm s}$ to this contrast in Fit 1 and less than $100\,{\rm s}$ in Fit 2, while the eccentricity-related contribution is only about $50\,{\rm s}$ in both fits. The relative phases of the geometric and apsidal contributions further affect the contrast observed in each epoch. These results illustrate how the changing disk-crossing geometry can produce a pronounced long-short pattern in the early epochs even for a low-eccentricity orbit.

\section{Summary and Discussion}

We have studied the secular evolution of eRO-QPE2 using 11 XMM-Newton observations spanning 4.33 years.
The eruption peak count rate remains approximately constant, while the quiescent count rate shows an overall decline with clear deviations from a linear trend.
The two components may come from different regions of the accretion disk, while disk precession may also contribute to the quiescent variations.
We find no evidence for secondary eruptions, with a 95\% upper limit of 0.56\% on their total counts relative to the normal eruptions.
Together with the tidal stability constraint on the star and the expected emission from disk ejecta and stellar debris, these results disfavor one eruption per period.

We then fitted the QPE timing with an EMRI and a precessing disk producing two eruptions per period, where the EMRI orbit is modeled as a forced trajectory rather than a geodesics in the Kerr spacetime. The fits to epochs 1-4 and 4-11 both favor a low-eccentricity orbit, with $e< 0.026$ and
$e<0.033$ at 95\% confidence, respectively. The two fits reveal consistent constraints on the orbital period evolution. The recurrence time shows a long-term decrease, with an average rate of $\dot{T}_{{\rm rec}}\simeq\frac{1}{2}\dot{T}_{{\rm obt}}\approx-5\times10^{-6}$. No clear evidence for evolution of disk parameters is identified.

There are two interesting implications if the EMRI+TDE disk interpretation of eRO-QPE2 is correct. The overall decay trend in the quiescent emission and the stable eruption emission imply a 
TDE disk with a decreasing accretion rate in the inner region where the quiescent X-ray emission comes from and  a stable surface density at $r=\mathcal{O}(10^2)M_\bullet$ where the EMRI crosses the disk.
As the disk precesses, the eruption emission remains approximately constant, which implies the EMRI incident angle does not affect the eruption energy much.
Both implications may  not be  naively expected.

\section*{Acknowledgements}

The computations in this paper were run on the Astro cluster supported by the Astronomy and Astrophysics Division of Tsung-Dao Lee Institute, Shanghai Jiao Tong University. 

\software{\texttt{SAS} \citep{gabriel_sas_2004}, \texttt{bilby} \citep{ashton_bilby_2019}, \texttt{dynesty} \citep{speagle_dynesty_2020}.}

\begin{figure}[hp]
\includegraphics[width=1\columnwidth,keepaspectratio]{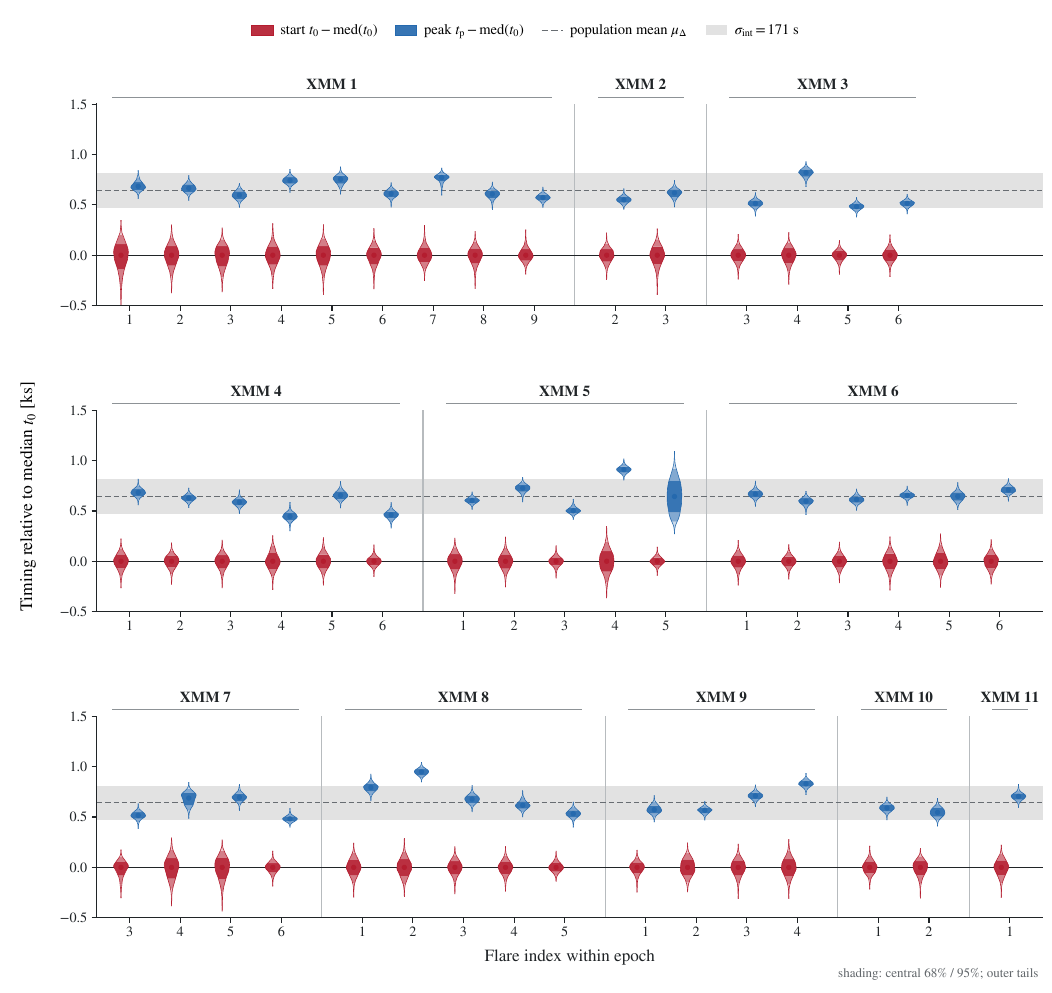} \caption{Violin plot of $t_{p}-t_{0}$ for the selected eRO-QPE2 flares. \textbf{Red:} flare starting time $t_{0}$. \textbf{Blue:} flare peak time $t_{p}$. The start time is fixed as the reference for each flare. The gray dashed line shows the mean of $t_{p}-t_{0}$, which is $628\,\mathrm{s}$, with intrinsic standard deviation $\sigma_{{\rm int}}=171^{+23}_{-17}\,\mathrm{s}$.}
\label{fig:timing-offset} 
\end{figure}

\begin{figure}
\includegraphics[width=1\columnwidth,keepaspectratio]{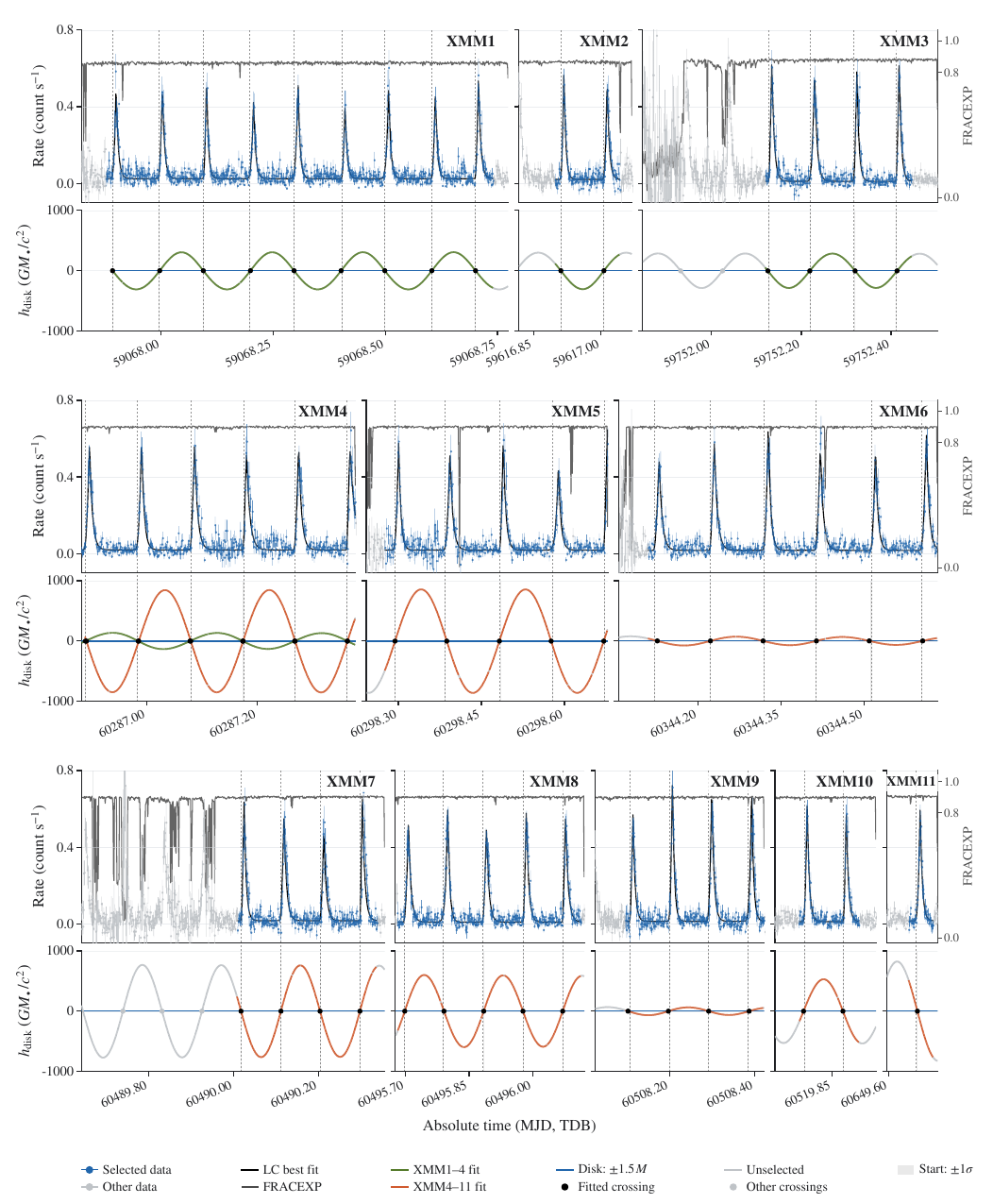}

\caption{eRO-QPE2 data and its phenomenological model fit (upper panels) together with best fit of flare timing model (lower panels). \textbf{Blue points:} data points of eRO-QPE2 in the $0.2\sim2\,\mathrm{keV}$ energy band with $100\,\mathrm{s}$ bin. \textbf{Blue curve:} light curve fitted with our phenomenological model. \textbf{Black points:} the fraction of effective exposure time (FRACEXP) within each time bin relative to the total duration of that bin. \textbf{Gray points:} points not selected because of low FRACEXP or because they do not include the start of an eruption. \textbf{Green \& orange lines:} distance between the orbiting secondary in the EMRI and the disk plane under our best fits for XMM1-4 and 4-11; their intersections with the blue line (corresponding disk plane) denote the QPE flare timings.}
\label{fig:timing-fits} 
\end{figure}

\appendix
Upper panels of Fig.~\ref{fig:timing-fits} show the eRO-QPE2 light curves for all 11 XMM epochs together with their fits with a phenomenological model \citep{Norris:2005ew,Arcodia+2022}. The flare peak time $t_{p}$ and the flare start time $t_{0}$ (defined as when the flare luminosity first rises to $1/e^{3}$ times its peak value) are commonly used in QPE timing analyses. If all flares have identical profiles and constant widths, the two definitions are equivalent. Otherwise, variations in the flare profile introduce additional QPE timing uncertainties. We analyze all the flares of eRO-QPE2 and find that $t_{p}-t_{0}$ varies between flares, with a mean of $628~\mathrm{s}$ and an intrinsic standard deviation of $\sigma_{{\rm int}}=171^{+23}_{-17}~\mathrm{s}$ (Fig.~\ref{fig:timing-offset}). This intrinsic scatter provides a measure of the timing uncertainty caused by flare profile variations.

\begin{figure*}
\begin{centering}
\includegraphics[width=1\textwidth,keepaspectratio]{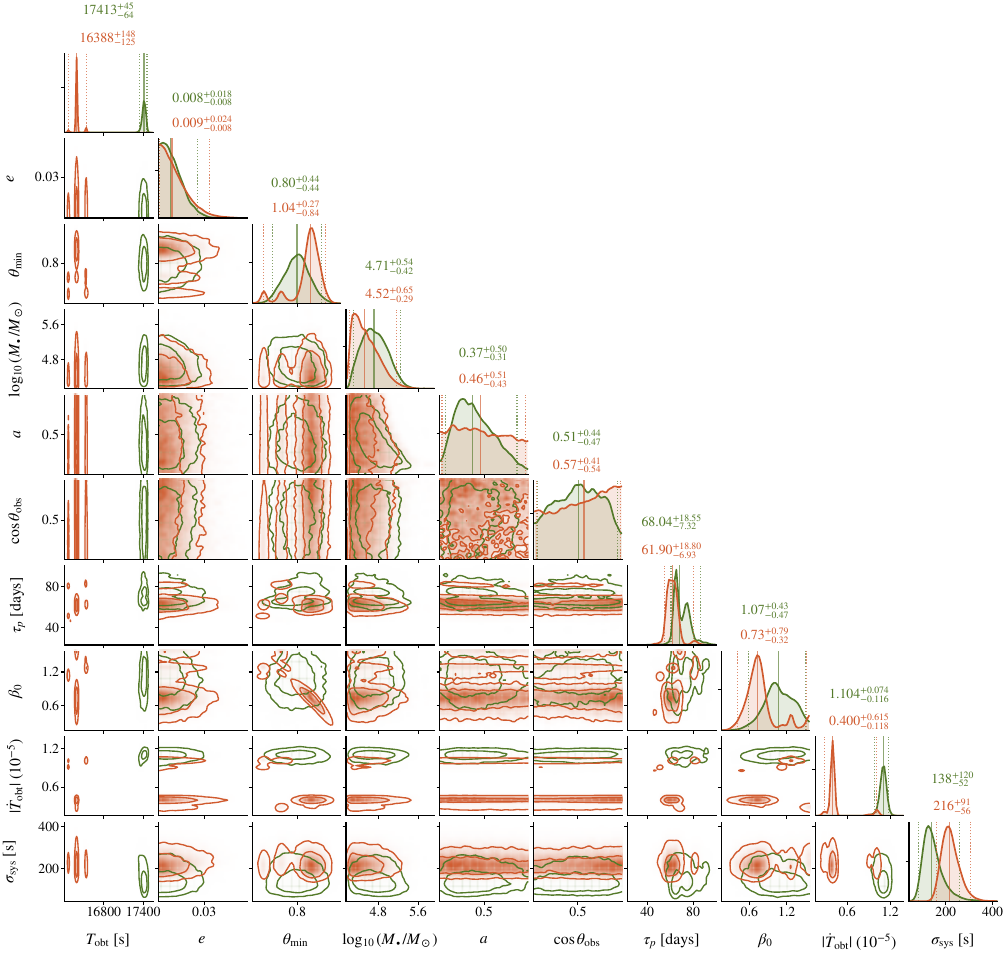} 
\par\end{centering}
\caption{Posterior distributions of the non-phase parameters from the two fits. \textbf{Green:} epochs $1$-$4$. \textbf{Orange:} epochs $4$-$11$. The initial orbital periods refer to the first observed flare in each interval. The annotations on the diagonal give the posterior medians and their $95\%$ credible intervals.}
\label{fig:corner} 
\end{figure*}

In this work, we use the flare starting time $t_0$ in QPE timing analyses, where we introduce an extra parameter $\sigma_{\rm sys}$ taking unmodeled physical processes including the intrinsic flare profile variations into account
(see Eq.~(\ref{eq:likelihood})).
Lower panels of Fig.~\ref{fig:timing-fits} show the best-fit EMRI+disk model of epochs 1-4 and epochs 4-11, respectively.  The full posteriors of model parameters are shown in Fig.~\ref{fig:corner}, except the phase parameters which are loosely bounded and not informative.

\clearpage{}

\bibliography{qpe_references}

\end{document}